\documentclass[aip,jmp,amsmath,amssymb,reprint]{revtex4-1}

\usepackage{graphicx}
\usepackage{dcolumn}
\usepackage{bm}

\begin{document}

\preprint{AIP/123-QED}

\title{Giant magnetocaloric effect in magnetically frustrated EuHo$_2$O$_4$ and EuDy$_2$O$_4$ compounds}

\author{A. Midya, N. Khan, D. Bhoi and P. Mandal}
\email{prabhat.mandal@saha.ac.in}
\affiliation{Saha Institute of Nuclear Physics, 1/AF Bidhannagar, Calcutta 700 064, India}
\date{\today}

\begin{abstract}
We have investigated the magnetic and magnetocaloric properties of EuHo$_2$O$_4$ and EuDy$_2$O$_4$  by magnetization and heat capacity measurements down to 2 K. These compounds undergo a field-induced  antiferromagnetic  to ferromagnetic  transition and exhibit a huge entropy change. For a field change of 0-8 T, the maximum magnetic entropy  and  adiabatic temperature changes are 30 (25)  J kg$^{-1}$ K$^{-1}$ and 12.7 (16) K, respectively and  the corresponding value of refrigerant capacity  is 540 (415) J kg$^{-1}$ for EuHo$_2$O$_4$ (EuDy$_2$O$_4$). These magnetocaloric parameters also remain   large  down to lowest temperature  measured and are even larger than that for some of the potential magnetic refrigerants  reported in the same temperature range for a moderate field change. Moreover, these materials are highly insulating  and exhibit no thermal and field  hysteresis, fulfilling the necessary conditions for a good magnetic refrigerant in the low-temperature region.
\end{abstract}


\maketitle

Research on magnetic refrigeration based on magnetocaloric effect (MCE) has received considerable attention  for their energy efficiency and elimination of environmentally harmful chlorofluorocarbon  gas which is used in a conventional vapor cycle refrigeration\cite{kag}. The parameter which describes the magnetocaloric effect is the magnetic entropy change ($\Delta S_M$) in an adiabatic process under external magnetic field\cite{kag,tishin}. Large  MCE in the low-temperature region would be useful for some specific technological applications such as space science, liquefaction of hydrogen in fuel industry while the large MCE close to room temperature can be used for domestic and industrial refrigerant purposes\cite{kag,prov,bfy}. The materials which  exhibit a large entropy change at the ferromagnetic (FM) to paramagnetic (PM) transition or field-induced metamagnetic transition from antiferromagnetic (AFM) to FM state with a minimal hysteresis having a low heat capacity are the potential candidates for technological applications. The magnetic entropy change can be large for  the field-induced first-order phase transition in which magnetic and structural phases are  coupled  or in a metamagnetic transition. However, due to the thermal and field hysteresis of the first-order phase transition, the refrigerant capacity  of the material is reduced. Often, materials showing field-induced AFM-FM transition exhibit huge magnetic entropy change without any thermal and field hysteresis. \\

Ternary compounds  Eu$Ln_2$O$_4$ ($Ln$$=$Gd-Yb) crystallize in the orthorhombic CaFe$_2$O$_4$ structure in which the lanthanide ions are forming zigzag chains with a honeycomb-like structure\cite{his,hol}. In these geometrically frustrated magnetic materials, a large number of different ground states have been observed  which is an active area of experimental and theoretical  research. It has been observed that the susceptibility of similar compounds, Sr$Ln_2$O$_4$\cite{cava}  and Ba$Ln_2$O$_4$,\cite{doi}  show an anomaly, which is ascribed to the magnetic interaction between the $Ln^{3+}$ ions because the alkali ions, Sr$^{2+}$ and Ba$^{2+}$,  are nonmagnetic. By contrast, the Eu$^{2+}$ ions at the alkali site in Eu$Ln_2$O$_4$ are expected to introduce additional magnetic interactions with the $Ln^{3+}$ ions, thus affecting the magnetic behavior due to their large magnetic moment arising from partially occupied 4f orbital. As the magnetic entropy  depends on the total angular momentum $J$, the introduction of Eu at Sr site increases the total angular momentum and, therefore, one  expects a large entropy change near the magnetic transition in Eu$Ln_2$O$_4$. Here, we present the  magnetic and magnetocaloric properties of EuHo$_2$O$_4$ and EuDy$_2$O$_4$ materials. As both Ho and Dy ions have large angular momentum,  a large entropy change is expected to occur with applied field.  Indeed, our results demonstrate that these compounds are suitable for magnetic refrigerant in the low-temperature region due to their giant MCE, large adiabatic temperature change, and  large relative cooling power (RCP).\\

We have prepared the polycrystalline EuHo$_2$O$_4$ and EuDy$_2$O$_4$ samples by solid state reaction method. High purity Eu$_2$O$_3$, Ho$_2$O$_3$/Dy$_2$O$_3$ and Dy/Ho were mixed in appropriate ratios. The mixture was then heated in an evacuated quartz tube at 1000 $^{\circ}$C for 30 h. Finally, the samples were prepared by heating in a quartz tube at 1100 $^{\circ}$C for 30 h with an intermediate grinding in argon atmosphere. The structural  analysis was performed by using powder x-ray diffraction technique (Rigaku, TTRAX II) and the results are consistent with those in a previous report\cite{his}.  The temperature and field dependent dc magnetization ($M$) and zero-field heat capacity ($C_p$) were measured in a physical properties measurement system (Quantum Design). \\
\begin{figure}[]
\includegraphics[height=11cm]{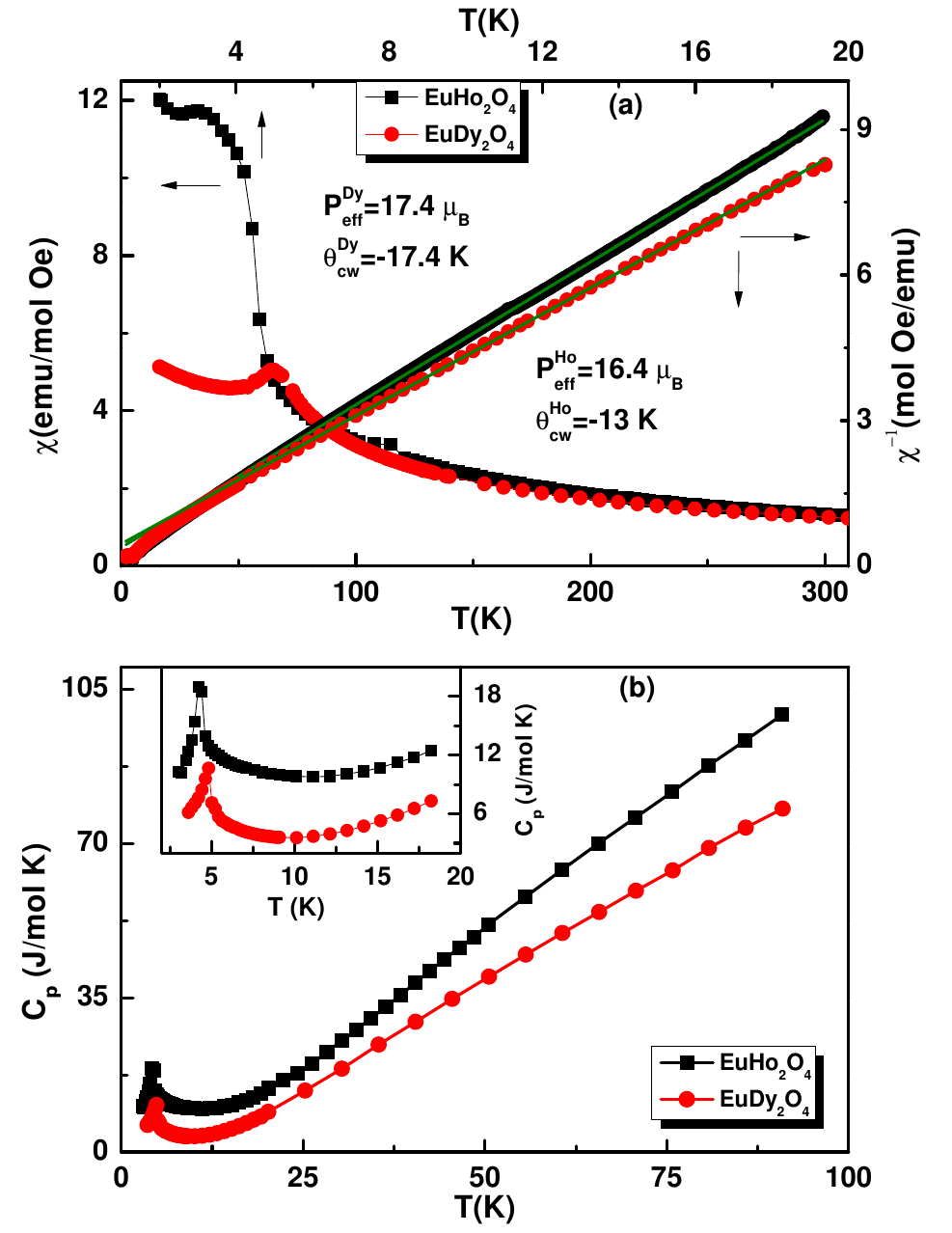}
\caption{
\label{fig:1} Fig. 1: (a) Temperature dependence of the field-cool dc susceptibility $\chi$ ($=$$M$/$H$) for $H$$=$ 100 Oe for EuHo$_2$O$_4$ and EuDy$_2$O$_4$. The right axis shows $\chi ^{-1}$($T$)  and the corresponding Curie-Weiss fit (solid line). (b) Temperature dependence of the zero-field specific heat for both the compounds.}
\end{figure}
The isothermal magnetic entropy change $\Delta S_M $ with field variation is given by $\Delta S_M(T,\Delta H)=\int_{H_i}^{H_f} \frac{\partial M}{\partial T} dH$. As the magnetization measurements were performed using desecrate temperature and magnetic field intervals, $\Delta S_M(T,\Delta H)$ has been estimated numerically by approximating the above equation as
\begin{equation}
\Delta S_{M}(T,H) = \sum_{i}\frac{M_{i+1} - M_{i}}{T_{i+1} - T_{i}}H_{i},
\label{eq1}
\end{equation}
where $M_{i}$ and $M_{i+1}$ are the experimentally measured values of magnetization for a magnetic field $H_i$  at temperatures $T_i$ and $T_{i+1}$, respectively. The refrigerant capacity or relative cooling power  is an important quality factor of the refrigerant material which is a measure of the amount of heat transfer between the cold and hot reservoirs in an ideal refrigeration cycle and is defined as, $RCP=\int\limits_{T_1}^{T_2}\Delta S_{M}dT$, where $T_1$ and $T_2$ are the temperatures corresponding to both sides of the half-maximum value of $\Delta S_{M}$($T$) peak. The adiabatic temperature change $\Delta T_{ad}$,  the another important factor related to magnetic refrigeration, is the isentropic temperature difference between $S(0,T)$ and $S(H,T)$. $\Delta T_{ad}$ may be calculated from the field-dependent magnetization and zero-field heat capacity data. $S(H,T)$ can be evaluated by subtracting the corresponding $\Delta S_M$ from $S(0,T)$, where the total entropy S(0,T) in absence of magnetic field is given by, $S(0, T)=\int\limits_{0}^{T}\frac{C_p(0, T)}{T}dT$.\\

The thermal evolution of  zero-field-cool (ZFC) and field-cool (FC) dc susceptibility $\chi$ ($=$$M/H$) have been measured at $100$ Oe for  both EuHo$_2$O$_4$ and EuDy$_2$O$_4$.  No significant difference between ZFC and FC cycles has been observed in $\chi$. Figure 1(a) shows the temperature dependence of field-cool $\chi$. For EuDy$_2$O$_4$, $\chi$($T$)  shows a peak at around $T_N$$=$5 K which is a characteristic of magnetic transition from AFM to PM states. However, the nature of $\chi$($T$) at low temperature for EuHo$_2$O$_4$ compound is very different from that for EuDy$_2$O$_4$. With the decrease of $T$, $\chi$ increases abruptly at around 5 K and then passes through a broad maximum at around 3 K. With further decrease of $T$ below 2.5 K, $\chi$ increases very slowly. This behavior signifies that in EuHo$_2$O$_4$ neither AFM nor FM interaction is dominating but both the interactions are of comparable strength.  It may be mentioned here that in EuDy$_2$O$_4$ too, the peak due to AFM transition disappears and the nature of $T$ dependence of $\chi$  at low temperatures is qualitatively similar to  that for EuHo$_2$O$_4$ when the applied field exceeds only few hundreds Oe. This suggests that the AFM interaction in EuDy$_2$O$_4$ is also very weak. We will discuss this issue in more details in the later section. In the PM state, $\chi$ for both the compounds  show similar $T$ dependence; $\chi$  obeys the Curie-Weiss (CW) law [$\chi$$=$$C/(T+\theta)$].  From the linear fit of inverse of $\chi$, we have calculated the effective magnetic moment $P_{eff}$$=$$17.4$ $\mu_B$ and CW temperature $\theta$$=-$17.4 K for EuDy$_2$O$_4$ and the corresponding values are $16.5$ $\mu_B$ and -13 K for EuHo$_2$O$_4$. The observed $P_{eff}$ is close to the theoretically expected moment, calculated using the two-sublattice model $P_{eff}$=$\sqrt{ (P_{eff}^{Eu})^2+(P_{eff}^{Ln})^2}$. The negative values of $\theta$  suggest a predominant FM  interaction between the nearest neighbor Eu$^{2+}$ moments within the chain and the FM chains are antiferromagnetically coupled, giving rise to an overall AFM structure.  Temperature dependence of specific heat shows a $\lambda$-like peak around $5$ K due to the magnetic ordering  as confirmed by the magnetization measurement [Fig. 1(b)].  \\
\begin{figure}[]
\includegraphics[height=10cm]{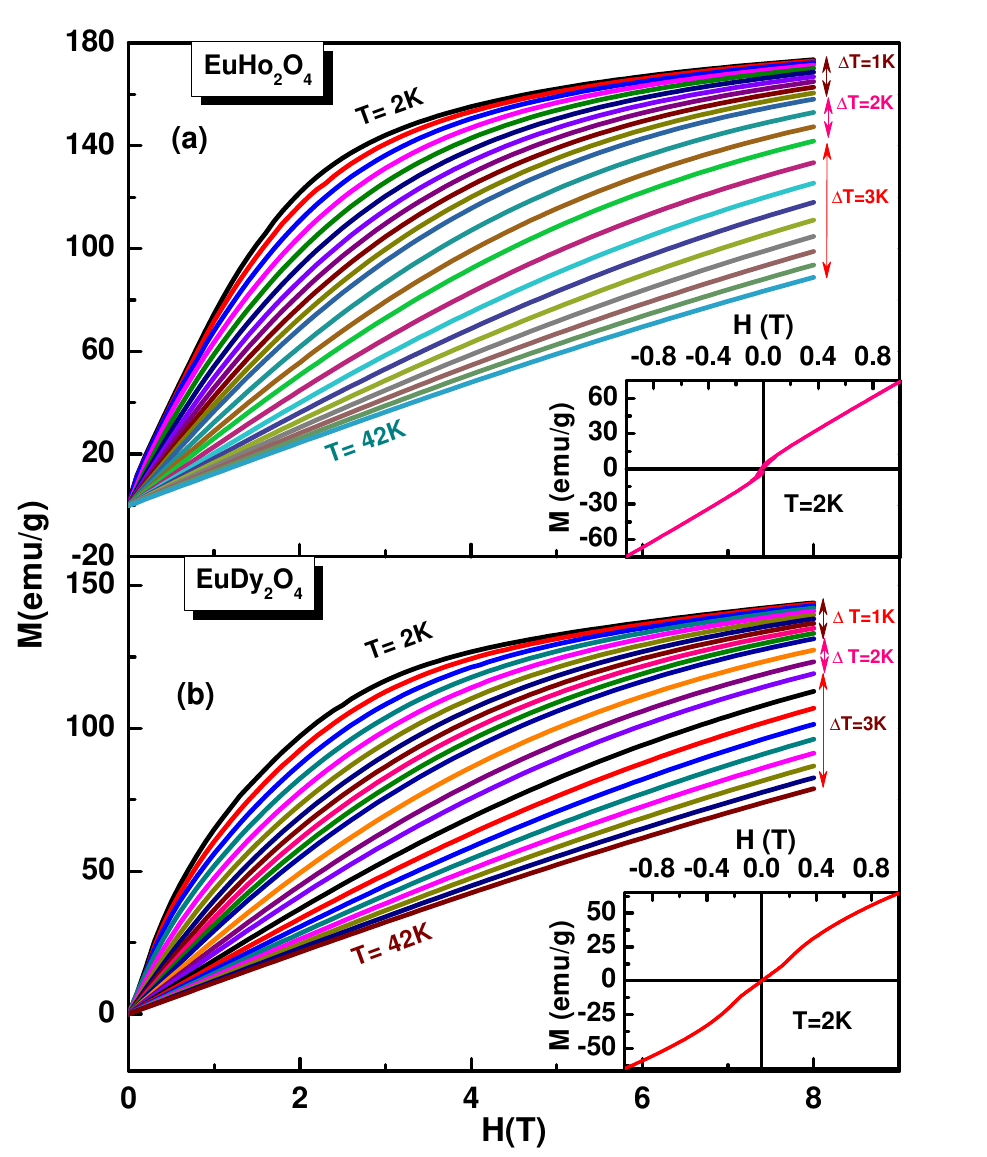}
\caption{
\label{fig:1}Isothermal magnetization for (a) EuHo$_2$O$_4$  and   (b) EuDy$_2$O$_4$ as a function of magnetic field for different temperatures. Insets show the low-field hysteresis at 2 K.}
\end{figure}

The isothermal $M$($H$)  curves at different temperatures  are shown in Fig. 2 for EuHo$_2$O$_4$ and EuDy$_2$O$_4$. For both the samples, $M$  increases smoothly with magnetic field. At low temperatures, though $M$ increases slowly with $H$ at high fields, no saturation-like behavior has been observed up to the highest applied magnetic field. For both the compounds, the observed values of magnetic moment at 2 K and 8 T are  substantially smaller than the local moments seen in the high temperature magnetic susceptibilities, indicative of the fact that all the spins cannot be aligned with the field up to 8 T. A qualitative similar behavior has been observed in Sr$Ln_2$O$_4$ compounds\cite{cava}.  The magnitude of magnetic moment increases monotonically with the decrease of temperature as in the case of a ferromagnet.  This  behavior suggests that the field-induced metamagnetic transition from AFM to FM state occurs  at a small value of applied field. The insets of Figs. 2(a) and 2(b) display the five-segment  $M$($H$) loop at 2 K up to 1 T. $M$($H$) does not show any hysteresis at low field.  In order to elucidate the nature of induced ferromagnetism in these compounds, we have also studied the temperature dependence of magnetization for different applied fields  (not shown). No thermal hysteresis between heating and cooling cycles of $M$ has been detected. We observe that $M$($T$) curves show a step-like behavior at temperatures above $T_N$ which corresponds to FM-PM transition. It may be noted that the field-induced FM transition temperature $T_C$ (defined as the position of the minimum in d$M$/d$T$ vs $T$ curve)  shifts to higher temperature continuously with increasing $H$ at the rate of 2 and 3 K/T for EuHo$_2$O$_4$ and EuDy$_2$O$_4$, respectively. \\
\begin{figure}[b]
\includegraphics[height=6cm]{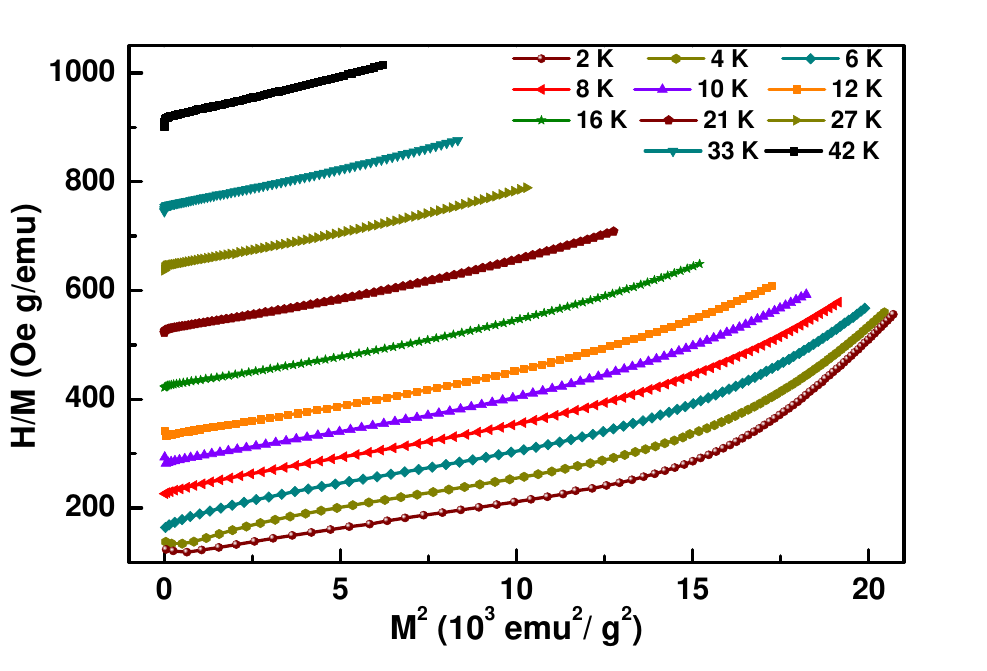}
\caption{
\label{fig:1} The Arrott plots for EuDy$_2$O$_4$ compound at some selected temperatures.}
\end{figure}
For further understanding the nature of field-induced magnetic transition, we have converted the $M$($H$) data in figure 2 into the Arrott plots\cite{arrot}. Figure 3 shows the Arrott plots  at different temperatures for EuDy$_2$O$_4$ compound. According to the Banerjee criterion \cite{sk}, a magnetic transition is expected to be of the first order when the slope of the Arrott plot is negative, whereas it will be of the second order when the slope is positive. The positive slope of the Arrott plots at low  as well as high fields implies that the field-induced FM transition above $T_N$ is second-order in nature. We have also done the Arrott plots for EuHo$_2$O$_4$ sample and the behavior is qualitatively similar to that for EuDy$_2$O$_4$ compound.\\

In order to test whether these materials  are suitable for magnetic refrigeration, we have calculated the isothermal magnetic entropy change using the Eq. 1. The temperature dependence  of $\Delta S_{M}$ for EuHo$_2$O$_4$ and EuDy$_2$O$_4$ are shown in figure 4 for different field variations up to 8 T.  $\Delta S_{M}$ is negative down to the lowest measured temperature  and the maximum value of $\Delta S_{M}$  ($\Delta S_{M}^{max}$) increases with field reaching  30 and 25 J kg$^{-1}$ K$^{-1}$ for a field change 0-8 T for EuHo$_2$O$_4$ and EuDy$_2$O$_4$, respectively. Also, the position of maximum in $\Delta S_{M}$($T$) curve shifts slowly toward higher $T$ with increasing $H$. It is clear from the figures that $\Delta S_{M}^{max}$ does not show saturation-like behavior even at high fields. Inset of Fig. 4 shows the variation of refrigerant capacity of the material with magnetic field. The maximum values of RCP for a field change of 8 T are 540 and 415 J kg$^{-1}$ for EuHo$_2$O$_4$ and EuDy$_2$O$_4$, respectively. Thus, both $\Delta S_{M}^{max}$ and RCP are quite large in these materials.  The large values of $\Delta S_{M}^{max}$ and RCP of the present compounds are comparable to those observed in several multiferroic manganites \cite{midya1,midya2}and ternary intermetallic compounds \cite{chen,li} but much larger than that observed in several perovskite manganites\cite{phan,guo} or Heusler alloys \cite{kren,sha}. The temperature dependence of adiabatic temperature change  for various magnetic fields are shown in Fig. 5. In EuDy$_2$O$_4$, the maximum value of $\Delta T_{ad}$  ($\Delta T_{ad}^{max}$) reaches as high as $16$ K for a field change of 8 T. From figures 4 and 5, it is clear that  the magnetocaloric parameters also have reasonably large value at a moderate field strength which is an important criterion for magnetic refrigeration.\\
\begin{figure}[]
\includegraphics[height=10cm]{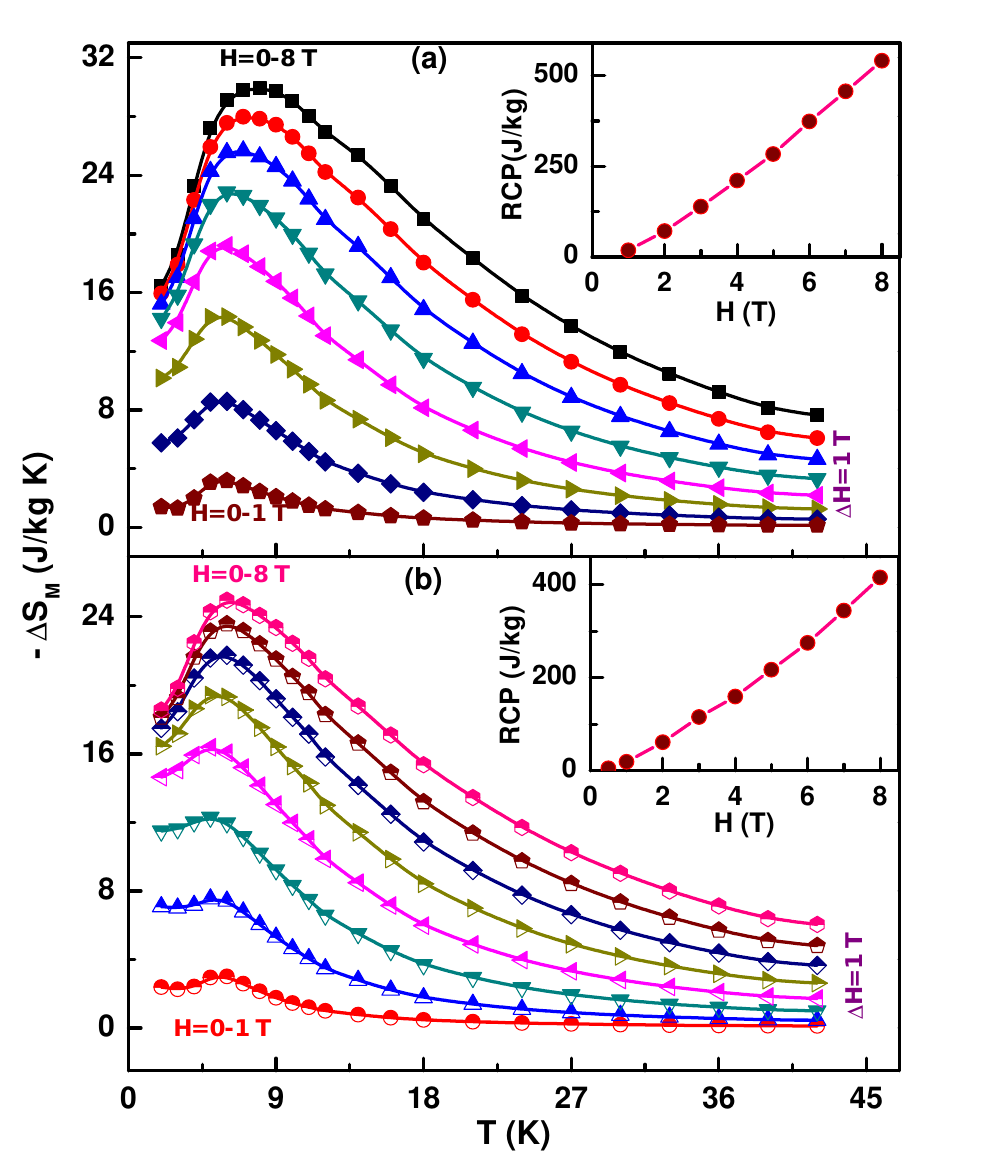}
\caption{
\label{fig:1} Temperature dependence of magnetic entropy change $\Delta S_{M}$ for (a) EuHo$_2$O$_4$  and  (b) EuDy$_2$O$_4$ compounds. Insets show the refrigerant capacity as a function of magnetic field.}
\end{figure}

\begin{figure}[b]
\includegraphics[height=6cm]{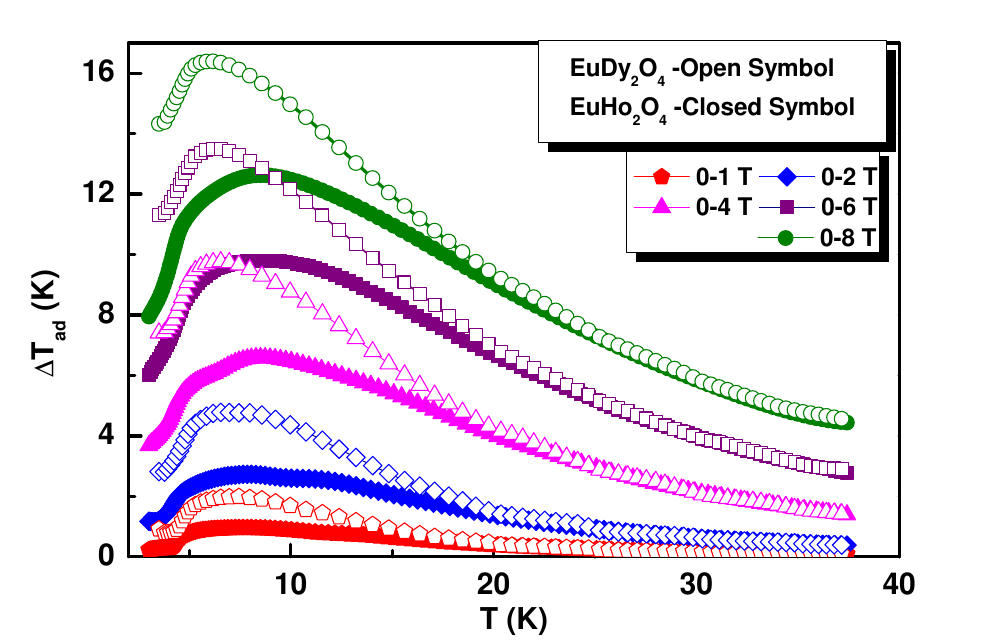}
\caption{
\label{fig:1}The adiabatic temperature change ($\Delta T_{ad}$) for EuHo$_2$O$_4$ (closed symbol) and  EuDy$_2$O$_4$ (open symbol)  as a function of temperature. }
\end{figure}
Both MCE and $T_{ad}$ have reasonably good pick-width and they do not drop abruptly to a small value well below $T_C$, indicating the high cooling efficiency even at very low temperature.  For example, in EuDy$_2$O$_4$, $\Delta S_{M}$ at 2 K is as high as 85 $\%$ of $\Delta S_{M}^{max}$ for the field change of 5 T. We have already mentioned that  several compounds exhibit large MCE,   RCP and  $\Delta T_{ad}$ as in the present case. However,  the  magnetocaloric parameters in these materials decrease rapidly below $T_C$ and, as a result, their cooling efficiency at low temperature is very poor. Normally, for a ferromagnetically ordered material, the distribution of $\Delta S_{M}$($T$) is highly asymmetric with respect to $\Delta S_{M}^{max}$. $\Delta S_{M}$($T$) exhibits a long tail in the PM state while it decreases rapidly at low temperatures below $T_C$ due to the saturation of $M$. However,  in the present compounds, the magnetization does not saturate at low temperatures even at a moderate field strength. We believe that this unusual behavior of $M$ aries due to the complicated low-dimensional magnetic structure and frustration. Structural, magnetic and neutron diffraction studies show that the magnetic sublattice of Sr$Ln_2$O$_4$ has several levels of low dimensionality and frustration, and the complexities of the resulting magnetic states at low temperatures vary from one lanthanide to another \cite{cava}. A more clearer picture  on the nature magnetic ground states emerges from the zero-field muon spin-relaxation studies on Eu$Ln_2$O$_4$ compounds \cite{ofer}. It has been shown that EuLu$_2$O$_4$ exhibits a static long-range AFM ordering below 5.7 K but when the nonmagnetic Lu$^{3+}$ is replaced by magnetic lanthanides then the long-range static ordering gets disrupted. For example, in EuGd$_2$O$_4$, the strong Gd moments destroy the local magnetic ordering and stabilize a dynamic disordered phase instead of static ordering.  As both Ho$^{3+}$ and Dy$^{3+}$ possess large magnetic moment like Gd$^{3+}$, one may expect a highly disordered magnetic ground state in EuHo$_2$O$_4$ and EuDy$_2$O$_4$ compounds similar to that observed in EuGd$_2$O$_4$. If it is so, then magnetization may not show the saturation-like behavior at low temperature as in the case of a typical ferromagnet and hence a large MCE at low temperatures well below $T_C$.\\

In summary, magnetic and magnetocaloric properties of EuHo$_2$O$_4$ and EuDy$_2$O$_4$ have been studied by magnetization and heat capacity measurements. These compounds exhibit field-induced metamagnetic transition from AFM to FM state which leads to a giant negative entropy change. The maximum values of $\Delta S_M$, $\Delta T_{ad}$ and  RCP are found to be   30  J kg$^{-1}$ K$^{-1}$, 13 K and 540 J kg$^{-1}$, respectively for EuHo$_2$O$_4$ while the corresponding values are 25  J kg$^{-1}$ K$^{-1}$, 16 K and 415 J kg$^{-1}$, respectively for EuDy$_2$O$_4$ for a field change of 0-8 T. The parameters $\Delta S_M$, $\Delta T_{ad}$ and  RCP also have reasonably good values for a moderate field change. Unlike several potential magnetic refrigerants with similar transition temperatures, the magnetocaloric parameters of these present compounds do not decrease abruptly at low temperatures well below $T_C$ owing to strong magnetic frustration. The excellent magnetocaloric properties of EuHo$_2$O$_4$ and EuDy$_2$O$_4$ compounds  make them attractive for active magnetic refrigeration down to very low temperature.\\

As the measured $M$ is much lower than the theoretically expected value, there is a scope for the further enhancement of saturation magnetization and hence MCE by adopting or changing the sample preparation technique.

\end{document}